\documentclass[adraft]{eptcs}
\providecommand{\event}{} 
\usepackage{underscore}           
\usepackage{bbm}

\usepackage{xcolor}

\usepackage{amsmath}
\usepackage{subcaption}

\usepackage{appendix}

\usepackage[utf8]{inputenc}
\usepackage{anyfontsize}
\usepackage{xspace}
\usepackage{float}

\usepackage[capitalize, noabbrev]{cleveref} 

\usepackage{wrapfig}
\usepackage{caption}

\usepackage{longtable}
\usepackage{tabu}

\usepackage{mathpartir}
\usepackage{bussproofs}

\usepackage{cmll} 
\usepackage{mathtools} 
\usepackage{stmaryrd} 
\usepackage{xfrac} 

\usepackage[qm,braket]{qcircuit}

\newcommand{\sqire}{s\textsc{qir}e\xspace}
\newcommand{\qwire}{\ensuremath{\mathcal{Q}\textsc{wire}}\xspace}

\newcommand{\CC}{\ensuremath{\mathbb{C}}\xspace}

\usepackage{tikz} 
\usetikzlibrary{%
  arrows,%
  shapes.misc,
  shapes.geometric, 
  shapes.arrows,%
  shapes.callouts,
  shapes.gates.logic.US,
  chains,%
  matrix,%
  positioning,
  scopes,%
  decorations.pathmorphing,
  decorations.text,
  decorations.pathreplacing, 
  shadows%
}
\tikzset{ machine/.style={
    rectangle,
    minimum width=25mm,
    minimum height=18mm,
    text width=24mm,
    align=center,
    very thick,
    draw=black,
    color=black,
    fill=white,
  }
}

\DeclarePairedDelimiter\abs{\lvert}{\rvert}
\DeclarePairedDelimiter\norm{\lVert}{\rVert}

\makeatletter
\let\oldabs\abs
\def\abs{\@ifstar{\oldabs}{\oldabs*}}
\let\oldnorm\norm
\def\norm{\@ifstar{\oldnorm}{\oldnorm*}}
\makeatother

\makeatletter
\DeclareRobustCommand{\vardivision}{%
  \mathbin{\mathpalette\@vardivision\relax}%
}
\newcommand{\@vardivision}[2]{%
  \reflectbox{$\m@th\smallsetminus$}%
}
\makeatother

\newcommand{\lbind}{\ensuremath{\mathbin{=\!\!>\!\!>\!\!=}}}

\usepackage{booktabs}

\usepackage{alltt}
\usepackage{listings,lstcoq}
\usepackage{MnSymbol}
\definecolor{ltblue}{rgb}{0,0.4,0.4}
\definecolor{dkblue}{rgb}{0,0.1,0.6}
\definecolor{dkgreen}{rgb}{0,0.35,0}
\definecolor{dkviolet}{rgb}{0.3,0,0.5}
\definecolor{dkred}{rgb}{0.5,0,0}
\newcommand{\code}[1]{{\small\texttt{#1}}}

\newcommand{\denote}[1]{\llbracket #1 \rrbracket\xspace}

\usepackage{newunicodechar}
\let\Alpha=A
\let\Beta=B
\let\Epsilon=E
\let\Zeta=Z
\let\Eta=H
\let\Iota=I
\let\Kappa=K
\let\Mu=M
\let\Nu=N
\let\Omicron=O
\let\omicron=o
\let\Rho=P
\let\Tau=T
\let\Chi=X

\newunicodechar{Α}{\ensuremath{\Alpha}}
\newunicodechar{α}{\ensuremath{\alpha}}
\newunicodechar{Β}{\ensuremath{\Beta}}
\newunicodechar{β}{\ensuremath{\beta}}
\newunicodechar{Γ}{\ensuremath{\Gamma}}
\newunicodechar{γ}{\ensuremath{\gamma}}
\newunicodechar{Δ}{\ensuremath{\Delta}}
\newunicodechar{δ}{\ensuremath{\delta}}
\newunicodechar{Ε}{\ensuremath{\Epsilon}}
\newunicodechar{ε}{\ensuremath{\epsilon}}
\newunicodechar{ϵ}{\ensuremath{\varepsilon}}
\newunicodechar{Ζ}{\ensuremath{\Zeta}}
\newunicodechar{ζ}{\ensuremath{\zeta}}
\newunicodechar{Η}{\ensuremath{\Eta}}
\newunicodechar{η}{\ensuremath{\eta}}
\newunicodechar{Θ}{\ensuremath{\Theta}}
\newunicodechar{θ}{\ensuremath{\theta}}
\newunicodechar{ϑ}{\ensuremath{\vartheta}}
\newunicodechar{Ι}{\ensuremath{\Iota}}
\newunicodechar{ι}{\ensuremath{\iota}}
\newunicodechar{Κ}{\ensuremath{\Kappa}}
\newunicodechar{κ}{\ensuremath{\kappa}}
\newunicodechar{Λ}{\ensuremath{\Lambda}}
\newunicodechar{λ}{\ensuremath{\lambda}}
\newunicodechar{Μ}{\ensuremath{\Mu}}
\newunicodechar{μ}{\ensuremath{\mu}}
\newunicodechar{Ν}{\ensuremath{\Nu}}
\newunicodechar{ν}{\ensuremath{\nu}}
\newunicodechar{Ξ}{\ensuremath{\Xi}}
\newunicodechar{ξ}{\ensuremath{\xi}}
\newunicodechar{Ο}{\ensuremath{\Omicron}}
\newunicodechar{ο}{\ensuremath{\omicron}}
\newunicodechar{Π}{\ensuremath{\Pi}}
\newunicodechar{π}{\ensuremath{\pi}}
\newunicodechar{ϖ}{\ensuremath{\varpi}}
\newunicodechar{Ρ}{\ensuremath{\Rho}}
\newunicodechar{ρ}{\ensuremath{\rho}}
\newunicodechar{ϱ}{\ensuremath{\varrho}}
\newunicodechar{Σ}{\ensuremath{\Sigma}}
\newunicodechar{σ}{\ensuremath{\sigma}}
\newunicodechar{ς}{\ensuremath{\varsigma}}
\newunicodechar{Τ}{\ensuremath{\Tau}}
\newunicodechar{τ}{\ensuremath{\tau}}
\newunicodechar{Υ}{\ensuremath{\Upsilon}}
\newunicodechar{υ}{\ensuremath{\upsilon}}
\newunicodechar{Φ}{\ensuremath{\Phi}}
\newunicodechar{φ}{\ensuremath{\phi}}
\newunicodechar{ϕ}{\ensuremath{\varphi}}
\newunicodechar{Χ}{\ensuremath{\Chi}}
\newunicodechar{χ}{\ensuremath{\chi}}
\newunicodechar{Ψ}{\ensuremath{\Psi}}
\newunicodechar{ψ}{\ensuremath{\psi}}
\newunicodechar{Ω}{\ensuremath{\Omega}}
\newunicodechar{ω}{\ensuremath{\omega}}

\newunicodechar{ℕ}{\ensuremath{\mathbb{N}}}
\newunicodechar{∅}{\ensuremath{\emptyset}}

\newunicodechar{∙}{\ensuremath{\bullet}}
\newunicodechar{≈}{\ensuremath{\approx}}
\newunicodechar{≅}{\ensuremath{\cong}}
\newunicodechar{≡}{\ensuremath{\equiv}}
\newunicodechar{≤}{\ensuremath{\le}}
\newunicodechar{≥}{\ensuremath{\ge}}
\newunicodechar{≠}{\ensuremath{\neq}}
\newunicodechar{∀}{\ensuremath{\forall}}
\newunicodechar{∃}{\ensuremath{\exists}}
\newunicodechar{±}{\ensuremath{\pm}}
\newunicodechar{∓}{\ensuremath{\pm}}
\newunicodechar{·}{\ensuremath{\cdot}}
\newunicodechar{⋯}{\ensuremath{\cdots}}
\newunicodechar{…}{\ensuremath{\ldots}}
\newunicodechar{∷}{~\mathrel{:\!\!\!:}~}
\newunicodechar{×}{\ensuremath{\times}}
\newunicodechar{∞}{\ensuremath{\infty}}
\newunicodechar{→}{\ensuremath{\to}}
\newunicodechar{←}{\ensuremath{\leftarrow}}
\newunicodechar{⇒}{\ensuremath{\Rightarrow}}
\newunicodechar{↦}{\ensuremath{\mapsto}}
\newunicodechar{↝}{\ensuremath{\leadsto}}
\newunicodechar{∨}{\ensuremath{\vee}}
\newunicodechar{∧}{\ensuremath{\wedge}}
\newunicodechar{⊢}{\ensuremath{\vdash}}
\newunicodechar{⊣}{\ensuremath{\dashv}}
\newunicodechar{∣}{\ensuremath{\mid}}
\newunicodechar{∈}{\ensuremath{\in}}
\newunicodechar{⊆}{\ensuremath{\subseteq}}
\newunicodechar{⊂}{\ensuremath{\subset}}
\newunicodechar{∪}{\ensuremath{\cup}}
\newunicodechar{⋓}{\ensuremath{\Cup}}
\newunicodechar{∉}{\ensuremath{\not\in}}
\newunicodechar{√}{\ensuremath{\sqrt}}

\newunicodechar{⊸}{\ensuremath{\multimap}}
\newunicodechar{⊗}{\ensuremath{\otimes}}
\newunicodechar{⨂}{\ensuremath{\bigotimes}}
\newunicodechar{⊕}{\ensuremath{\oplus}}
\newunicodechar{〈}{\ensuremath{\langle}}
\newunicodechar{⟨}{\ensuremath{\langle}}
\newunicodechar{⟩}{\ensuremath{\rangle}}
\newunicodechar{〉}{\ensuremath{\rangle}}
\newunicodechar{¡}{\ensuremath{\upsidedownbang}}
\newunicodechar{∘}{\ensuremath{\circ}}
\newunicodechar{†}{\ensuremath{\dagger}}
\newunicodechar{⊤}{\ensuremath{\top}}
\newunicodechar{⊥}{\ensuremath{\bot}}

\newunicodechar{〚}{\ensuremath{\llbracket}}
\newunicodechar{〛}{\ensuremath{\rrbracket}}

\usepackage{etoolbox} 
\newtoggle{comments}
\toggletrue{comments}

\iftoggle{comments}{
  \newcommand{\fixme}[1]{\textbf{\textcolor{red}{[ Fixme: #1]}}}
  \newcommand{\todo}[1]{\textbf{\textcolor{green}{[ TODO: #1 ]}}}
  \newcommand{\rnr}[1]{\textbf{\textcolor{blue}{[ Robert: #1 ]}}}
  \newcommand{\mwh}[1]{\textbf{\textcolor{green}{[ Mike: #1 ]}}}
  \newcommand{\khh}[1]{\textbf{\textcolor{orange}{[ Kesha: #1 ]}}}
  \newcommand{\shh}[1]{\textbf{\textcolor{purple}{[ Shih-Han: #1 ]}}}
  \newcommand{\xwu}[1]{\textbf{\textcolor{purple}{[ Xiaodi: #1 ]}}}
}{
  \newcommand{\fixme}[1]{}
  \newcommand{\todo}[1]{}
  \newcommand{\rnr}[1]{}
  \newcommand{\mwh}[1]{}  
  \newcommand{\khh}[1]{}
  \newcommand{\shh}[1]{}
  \newcommand{\xwu}[1]{}
}

\title{Verified Optimization in a Quantum Intermediate Representation}
 \author{Kesha Hietala \qquad Robert Rand \qquad Shih-Han Hung
\and
Xiaodi Wu \qquad Michael Hicks
\institute{University of Maryland, College Park, USA}
\email{\{kesha, rrand, shung, xwu, mwh\}@cs.umd.edu}
}
\def\titlerunning{Verified Optimization in a Quantum Intermediate Representation}
\def\authorrunning{K. Hietala, R. Rand, S. Hung, X. Wu \& M. Hicks}
\begin{document}
\maketitle

\begin{abstract}

We present \sqire, a low-level language for expressing and formally verifying quantum programs. \sqire uses a global register of quantum bits. Doing so allows easy compilation to and from existing ``quantum assembly'' languages and simplifies the verification process. We demonstrate the power of \sqire as a compiler intermediate representation of quantum programs by verifying a number of useful optimizations, and we demonstrate \sqire's use as a tool for general verification by proving several quantum programs correct.

\end{abstract}

\section{Introduction}

Programming quantum computers, at least in the near term, will be challenging. Qubits will be scarce, and the risk of decoherence means that gate pipelines will need to be short. These limitations encourage the development of sophisticated algorithms and clever optimizations that are likely to have mistakes. For example, Nam et al.~\cite{Nam2018} discovered mistakes in both the theory and implementation of optimizing circuit transformations they developed, and found that the optimization library they compared against sometimes produced incorrect results. Unfortunately, we cannot apply standard software assurance techniques to address these challenges: Unit testing and debugging are infeasible due to both the indeterminacy of quantum algorithms and the substantial expense involved in executing or simulating them. 

To address these challenges, we can apply rigorous \emph{formal methods} to the development of quantum programs and programming languages. These methods aim to ensure mathematically-proved correctness of code by construction. A notable success of formal methods for classical computing is CompCert~\cite{compcert}, a \emph{certified compiler} for C programs. CompCert is written and \emph{proved correct} using the Coq proof assistant~\cite{coq}. CompCert includes sophisticated optimizations whose proofs of correctness are verified to be valid by Coq's type checker. An experimental evaluation of CompCert's reliability provided strong evidence of the validity of Coq's proof checking: While bug finding tools found hundreds of defects in the gcc and LLVM C compilers, no bugs were found in CompCert's verified core~\cite{Yang2011}.

As a first step toward a certified compiler for quantum programs, Rand et al.~\cite{Rand2018} developed a proved-correct compiler, written in Coq, from a source language describing Boolean functions to reversible oracles expressed as programs in the \qwire quantum circuit language~\cite{Paykin2017, RandThesis, Rand2017}. \qwire imposes well-formedness constraints on programs to ensure that resources are used properly, e.g., that all qubits are measured exactly once, and that ancillae are returned to their original state by the end of the circuit. \qwire programs are given a mathematical semantics in terms of density matrices, which is the foundation of proofs of correctness.

A certified compiler should optimize a quantum circuit and map it to machine resources. Rather than express these steps in \qwire, e.g., as source-to-source transformations, this paper proposes that they should be carried out in a simpler quantum language we call \sqire (pronounced ``squire''). While \qwire treats wires abstractly as Coq variables via higher order abstract syntax~\cite{Pfenning1988}, \sqire accesses qubits via concrete indices into a global register (\cref{sec:sqire,sec:general-sqire}). 
\sqire's simple design sacrifices some desirable features of high-level languages, such as variable binding to support easy compositionality.
This is not a significant drawback when using the language for intermediate-level programs, produced as the output of compiling a higher-level language.
Furthermore, low-level languages (like  OpenQASM \cite{Cross2017} and QUIL \cite{Smith2016}, both similar to \sqire) are more practical for programming near-term devices, which must be aware of both the number of qubits available and their connectivity~\cite{Preskill2018}. We discussed the challenges of programming and verifying such devices in a recent position paper \cite{Rand2019}.

As a demonstration of \sqire's utility as an intermediate representation, we have written several optimizations/transformations of \sqire programs that we have proved correct (Section~\ref{sec:optimizations}): \emph{Skip elimination}, \emph{Not propagation}, and \emph{Circuit layout mapping}. These transformations were prohibitively difficult to prove in \qwire but were relatively straightforward in \sqire. 

We find that \sqire is not only useful as the target of compilation and optimization. Its simple structure and semantics assists in proving correctness properties about programs written in \sqire directly. In particular, we have proved that the \sqire program to prepare the GHZ state indeed produces the correct state, and showed the correctness of quantum teleportation and the Deutsch-Jozsa algorithm (\cref{sec:general-verification}). 

The problem of quantum program optimization verification has previously been considered in the context of the ZX calculus \cite{Fagan2018}, but, as far as we are aware, our \sqire-based transformations are the first certified-correct optimizations applied to a realistic quantum circuit language.
Amy et al.~\cite{amy18reversible} developed a proved-correct optimizing compiler from source Boolean expressions to reversible circuits, but did not handle general quantum programs. Rand et al.~\cite{Rand2018} developed a similar compiler for quantum circuits but without optimizations. Prior low-level quantum languages \cite{Cross2017, Smith2016} have not been developed with verification in mind, and prior circuit-level optimizations \cite{Amy2013, Heyfron2017, Nam2018} have not been formally verified. Some recent efforts have examined using formal methods to prove properties of quantum computing source programs, e.g., Quantum Hoare Logic \cite{Ying2011}. This line of work is complementary to ours---a property proved of a source program is provably preserved by a certified compiler.
In addition, \sqire can also be used to prove properties about quantum programs by reasoning directly about their semantics.

Our work on \sqire constitutes a step toward developing a full-scale verified compiler toolchain. Next steps include developing certified transformations from high-level quantum languages to \sqire and implementing more interesting program transformations. We also hope that \sqire will prove useful for teaching concepts of quantum computing and verification in the style of the popular Software Foundations textbook~\cite{Pierce2016}.

All code we reference in this paper can be found at \url{https://github.com/inQWIRE/SQIRE}.

\section{\sqire: A Small Quantum Intermediate Representation}
\label{sec:sqire}

This section presents the syntax and semantics of \sqire programs.
To begin, we restrict our attention to the fragment of \sqire that describes unitary circuits.
We describe the full language, which allows measurement and initialization, in Section~\ref{sec:general-sqire}.

\subsection{Syntax and Semantics}

\begin{figure}[t]
\centering
\begin{subfigure}{.5\textwidth}
  \centering
  \begin{align*}
    P~\rightarrow&~skip \\
         &\vert~P_1 ;~P_2 \\
         &\vert~U~ q_1~...~q_n \\
         \\
    U~\rightarrow&~H~\vert~X~\vert~Y~\vert~Z~\vert~R_\phi~\vert~CNOT
  \end{align*}
\end{subfigure}%
\begin{subfigure}{.5\textwidth}
  \centering
  \begin{align*}
    \denote{skip}_u^{dim}&=I_{2^{dim}} \\
    \denote{P_1;~P_2}_u^{dim}&=~\denote{P_2}_u^{dim} \times \denote{P_1}_u^{dim} \\
    \denote{U~ q_1~...~q_n}_u^{dim}&=\begin{cases}
                          ueval(U,~q_1 ... q_n) &\text{well-typed} \\
                          0_{2^{dim}} &\text{otherwise}
                          \end{cases}
  \end{align*}
\end{subfigure}

  \caption{\sqire abstract syntax and semantics. We use the notation $\denote{P}_u^{dim}$ to describe the semantics of unitary program $P$ with a global register of size $dim$. $ueval(U,~q_1 ... q_n)$ returns the expected operation ($U$ for single-qubit gate $U$ and $\vert 1 \rangle \langle 1 \vert \otimes X + \vert 0 \rangle \langle 0 \vert \otimes I$ for $CNOT$), extended to the correct dimension by applying an identity operation on every other qubit in the system. For example, $ueval(H,~q) = I_{2^{q}} \otimes H \otimes I_{2^{dim - q - 1}}$.}
  \label{fig:sqire-syntax-semantics}
\end{figure}

\sqire is a low-level language primarily designed to be used as an intermediate representation in compilers for quantum programming languages. 
It is built on top of the Coq libraries developed for the \qwire language.
The main simplification in \sqire, compared to \qwire, is that it assumes a global register of qubits.
In \sqire, a qubit is referred to by a natural number that indexes into the global register whereas in \qwire qubits are referred to using standard Coq variables through the use of higher-order abstract syntax~\cite{Pfenning1988}.
The benefits and drawbacks of this simplification are discussed in Section~\ref{sec:global-register}.

Unitary \sqire programs allow three operations: skip, sequencing, and unitary application (of a fixed set of gates), as shown on the left of Figure~\ref{fig:sqire-syntax-semantics}.
Unitary application takes a list of indices into the global register.
A unitary program is well-typed if every unitary is applied to valid arguments. 
A list of arguments is valid if the length of the list is equal to the arity of the unitary operator, every element in the list is bounded by the dimension of the global register, and every element of the list is unique.
The first two properties ensure standard well-formedness conditions (function arity and index bounds) while the third enforces linearity and thereby quantum mechanics' no-cloning theorem. 
The Coq definitions of unitary \sqire programs and well-typedness are shown in \cref{fig:coq-sqire-definition}.

\begin{figure}[t]

\begin{tabular}{c}
\begin{coq}
Inductive ucom : Set :=
| uskip : ucom
| useq : ucom -> ucom -> ucom
| uapp : forall {n}, Unitary n -> list nat -> ucom.

Definition in_bounds (l : list nat) (max : nat) : Prop :=
    forall x, In x l -> x < max.

Inductive uc_well_typed : nat -> ucom -> Prop :=
| WT_uskip : forall dim, uc_well_typed dim uskip
| WT_seq : forall dim c1 c2, 
      uc_well_typed dim c1 -> uc_well_typed dim c2 -> uc_well_typed dim (c1; c2)
| WT_app : forall dim n l (u : Unitary n), 
      length l = n -> in_bounds l dim -> NoDup l -> uc_well_typed dim (uapp u l).
\end{coq}
\end{tabular}

    \caption{Coq definitions of unitary programs and well-typedness.}
    \label{fig:coq-sqire-definition}
\end{figure}

The semantics for unitary \sqire programs is shown on the right of Figure~\ref{fig:sqire-syntax-semantics}. 
If a program is well-typed, then we can compute its denotation in the expected way.
If a program is not well-typed, we ensure that its denotation is the zero matrix by returning zero whenever a unitary is applied to inappropriate arguments.
The advantage of this definition is that it allows us to talk about the denotation of a program without explicitly proving that the program is well-typed, which would result in proofs becoming cluttered with extra reasoning.

\sqire supports a fixed (universal) set of gates: $H$, $X$, $Y$, $Z$, $R_\phi$, and $CNOT$.
$R_\phi$ represents a phase shift by an arbitrary real number $\phi$.
In an effort to simplify the denotation function, the only multi-qubit gate we support is $CNOT$.
\sqire can be easily extended with other built-in gates, or new gates can be defined in terms of existing gates.
For example, we define the SWAP operation as follows.
\begin{coq}
Definition SWAP (a b : nat) : ucom := CNOT a b; CNOT b a; CNOT a b.
\end{coq}
We can then state and prove properties about the semantics of the defined operations. 
For example, we can prove that the SWAP program swaps its arguments, as intended.

\subsection{Example} \label{sec:sqire-example}

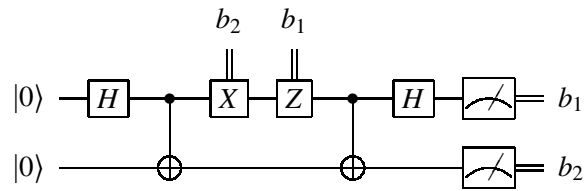
\begin{figure}[b]
\centerline{
  \Qcircuit @C=1em @R=1em {
     &  &  & b_2 & b_1 & & & & \\
     &  &  & \cwx[1] & \cwx[1] & & & & \\
    \lstick{\ket{0}} & \gate{H} & \ctrl{1} & \gate{X} & \gate{Z} & \ctrl{1} & \gate{H} & \meter & \rstick{b_1} \cw \\
    \lstick{\ket{0}} & \qw & \targ & \qw & \qw & \targ & \qw & \meter & \rstick{b_2} \cw
    }
}  
  \caption{Circuit for the superdense coding algorithm. }
  \label{fig:superdense-circ}
\end{figure}

Superdense coding is a protocol that allows a sender to transmit two classical bits, $b_1$ and $b_2$, to a receiver using a single quantum bit.
The circuit for superdense coding is shown in Figure~\ref{fig:superdense-circ}.
The \sqire program corresponding to the unitary part of this circuit is shown in Figure~\ref{fig:superdense-sqire}.
In the \sqire program, note that \texttt{encode} is a Coq function that takes two Boolean values and returns a circuit. 
This shows that although \sqire's design is simple, we can still express interesting quantum programs using help from the host language, Coq.
We will see additional examples of this style of metaprogramming in Section~\ref{sec:general-verification}.

\begin{figure}[t]
  \centering
    \begin{tabular}{c}
    \begin{coq}
Definition a : nat := 0.
Definition b : nat := 1.

Definition bell00 : ucom := H a; CNOT a b.

Definition encode (b1 b2 : bool): ucom :=
    (if b2 then X a else uskip);
    (if b1 then Z a else uskip).

Definition decode : ucom := CNOT a b; H a.

Definition superdense (b1 b2 : bool) := 
    bell00 ; encode b1 b2; decode.
    \end{coq}
    \end{tabular}

  \caption{\sqire program for the unitary portion of the superdense coding algorithm. Note that \coqe{U q} is syntactic sugar for applying unitary $U$ to qubit $q$.}
  \label{fig:superdense-sqire}

\end{figure}

Although \sqire was designed to be used as an intermediate representation, we can also prove properties about \sqire programs directly, since these programs and their semantics are embedded in Coq. For example, we can prove that the result of evaluating the program \coqe{superdense b1 b2} on an input state consisting of two qubits initialized to zero is the state $\vert b_1, b_2 \rangle$.
In our development, we write this as follows.
\begin{coq}
Lemma superdense_correct : forall b1 b2, 
    [[superdense b1 b2]]${}_u^2$ $\times$ $\vert$ 0,0 $\rangle$ = $\vert$ b1,b2 $\rangle$.
\end{coq}
Note that we are applying the denotation of \coqe{superdense} to a vector rather than a density matrix, and that we use Dirac (bra-ket) notation to represent this vector. 
In our experience, treating states as vectors and performing rewriting over bra-ket expressions simplifies reasoning.
With this is mind, we have added support for bra-ket reasoning to both \sqire and \qwire.

We will present additional examples of verifying correctness of \sqire programs in Section~\ref{sec:general-verification}.

\subsection{Discussion}
\label{sec:global-register}





The use of a global register significantly simplifies proofs about \sqire programs because register indices directly correspond to indices in the matrices that \sqire programs denote. By contrast, \qwire's variables map to different indices depending on the local context, which makes it difficult to make precise statements about program fragments. We elaborate on this issue in \cref{app:comparison}.

One downside of using a global register is that it does not allow for easy composition.
Combining separate \sqire programs requires manually defining a mapping from the global registers of both programs to a new, combined global register.
Furthermore, writing \sqire programs can be tedious because \sqire is a low-level language that references qubits only through natural numbers.
In contrast, writing \qwire programs is much like writing programs in any other high-level programming language. 
\qwire naturally allows composition of programs in a manner similar to normal function application. We further discuss the challenges of composition in \sqire in \cref{app:sqire-composition}.

We believe that \sqire's lower-level programming style, and the extra work required to perform composition, are not significant drawbacks in our use case. In particular, using \sqire as a compiler intermediate representation means that only the compiler has to deal with the extra details and tedium. Moreover, in our experience, it is not too difficult to write small \sqire programs manually. Such programs need not be built out of complicated, separate (and separately verified) parts, and do not involve managing many different qubits. The metalanguage can also ease composition and management of qubits (e.g., as done in \cref{fig:superdense-sqire}). We show more examples of directly expressed, and verified, \sqire programs in \cref{sec:general-verification}.

\section{General \sqire}
\label{sec:general-sqire}

To describe general quantum programs, we extend \sqire with operations for initialization and measurement. The command \coqe{meas q} measures a qubit and \coqe{reset q} measures a qubit and restores it to the $\ket{0}$ state.
We present two alternative semantics for general quantum programs: The first is based on density matrices and the second uses a non-deterministic definition to simplify reasoning. 

\subsection{Density Matrix Semantics}

Several previous efforts on verifying quantum programs have defined the semantics of quantum programs as operators on density matrices \cite{Paykin2017, Ying2011}. 
We follow this convention here, giving programs their standard interpretation.
The density matrix semantics of general \sqire programs is given in Figure~\ref{fig:sqire-syntax-semantics-general}. 

We prove the following correspondence between the density matrix semantics (denoted by subscript $d$) and the unitary semantics (subscript $u$) of \cref{fig:sqire-syntax-semantics}:
\begin{coq}
Lemma c_eval_ucom : forall (c : ucom) (dim : nat),
    [[c]]${}_d^{dim}$ = fun ρ => [[c]]${}_u^{dim}$ × ρ × ([[c]]${}_u^{dim}$)† .
\end{coq}
That is, the density matrix denotation of a unitary program simply multiplies the input state on both sides by the unitary denotation of the same program.

Note that our language does not include a construct for classical control (such as the ``if'' or ``while'' constructs in Ying's quantum while language \cite{Ying2011}). This is not a difficult extension, but we chose to keep \sqire simple to better reflect its intended use as a realistic intermediate representation for near-term quantum devices.

\begin{figure}[t]
\centering

  \centering
  \begin{align*}
    \denote{skip}_d^{dim}(\rho)&=I_{2^{dim}} \times \rho \times I_{2^{dim}}^\dagger \\
    \denote{P_1;~P_2}_d^{dim}(\rho)&=~(\denote{P_2}_d^{dim} \circ \denote{P_1}_d^{dim}) (\rho) \\
    \denote{U~ q_1~...~q_n}_d^{dim}(\rho)&=\begin{cases}
                          ueval(U) \times \rho \times ueval(U)^\dagger &\text{well-typed} \\
                          0_{2^{dim}} &\text{otherwise}
                          \end{cases} \\
   \denote{meas~q}_d^{dim}(\rho)&= \vert 0 \rangle_q \langle 0 \vert \rho \vert 0 \rangle_q \langle 0 \vert + \vert 1 \rangle_q \langle 1 \vert \rho \vert 1 \rangle_q \langle 1 \vert \\
   \denote{reset~q}_d^{dim}(\rho)&= \vert 0 \rangle_q \langle 0 \vert \rho \vert 0 \rangle_q \langle 0 \vert + \vert 0 \rangle_q \langle 1 \vert \rho \vert 1 \rangle_q \langle 0 \vert 
  \end{align*}

  \caption{\sqire density matrix semantics. We use the notation $\denote{P}_d^{dim}$ to describe the semantics of program $P$ with a global register of size $dim$. The definition of $ueval$ is given in the caption of Figure~\ref{fig:sqire-syntax-semantics}. We use $\vert i \rangle_q \langle j \vert$ as shorthand for $I_{2^q} \otimes \vert i \rangle \langle j \vert \otimes I_{2^{dim - q - 1}}$, which applies the projector to the relevant qubit and an identity operation to every other qubit in the system.}
  \label{fig:sqire-syntax-semantics-general}

\end{figure}

\subsection{Non-deterministic Semantics}

Our second semantics is the result of the observation that it is often useful, and simpler, to reason about quantum states as vectors rather than density matrices.
The non-deterministic semantics allows quantum states to be represented exclusively as vectors by allowing each outcome of a measurement to be reasoned about individually. An illustrative fragment of this semantics is given below.
\begin{coq}
Inductive nd_eval {dim : nat} : com -> Vector (2^dim) -> Vector (2^dim) -> Prop :=
  | nd_app : forall n (u : Unitary n) (l : list nat) (ψ : Vector (2^dim)),
      app u l / ψ ⇩ ((ueval dim u l) × ψ)
  | nd_meas0 : forall n (ψ : Vector (2^dim)),
      let ψ' := pad n dim \k0\b0 × ψ in 
      norm ψ' <> 0
      meas n / ψ ⇩ ψ' 
  | nd_meas1 : forall n (ψ : Vector (2^dim)),
      let ψ' := pad n dim \k1\b1 × ψ in
      norm ψ' <> 0 ->
      meas n / ψ ⇩ ψ' 

where "c '/' ψ '⇩' ψ'" := (nd_eval c ψ ψ').              
\end{coq}

Evaluation is given here as a relation.
The \coqe{nd_app} rule says that, given state $\psi$, \coqe{app u l} evaluates to \coqe{(ueval dim u l) $\times \psi$}, as expected.
The \coqe{nd_meas0} rule says that, if the result of projecting the $n^{th}$ qubit onto the $\ket{0}\bra{0}$ subspace is not the zero matrix, measuring $n$ yields this projection. 
Note that most quantum states can step via either the \coqe{nd_meas0} or \coqe{nd_meas1} rule. 
Reset behaves non-deterministically (like measurement) but sets the resulting qubit to $\ket{0}$.
To simplify the reasoning process, we do not rescale the output of measurement: As is standard in quantum computing proofs, the user may choose to reason about the normalized output of a program or to prove a property that is invariant to scaling factors.

We can show that the non-deterministic semantics of the unitary fragment of \sqire is identical to the unitary semantics, albeit in relational form:
\begin{coq}
Lemma nd_eval_ucom : forall (c : ucom) (dim : nat) (ψ ψ' : Vector (2^dim)),
    WF_Matrix ψ -> (c / ψ ⇩ ψ' <->  [[c]]${}_u^{dim}$× ψ = ψ').
\end{coq}
The \coqe{WF_Matrix} predicate here ensures that $\psi$ is a valid input to the circuit. 

We give an example of reasoning with both the density matrix semantics and the non-deterministic semantics in Section~\ref{sec:general-verification}.
The end goal of our work on this semantics, and other simplifications that we have made in the the design of \sqire, is to make \sqire a tool that can be used for intuitive reasoning by both teachers and practitioners.

\section{Verifying Program Transformations}
\label{sec:optimizations}

Because near-term quantum machines will only be able to perform small computations before decoherence takes effect, compilers for quantum programs must apply sophisticated optimizations to reduce resource usage. 
These optimizations can be complicated to implement and are vulnerable to programmer error.
It is thus important to verify that the implementations of program optimizations are correct.
Our work in this section is a first step toward a verified-correct optimizer for quantum programs.

We begin by discussing equivalence of \sqire programs.
We then discuss a simple optimization on unitary programs that removes all possible skip gates. 
We follow this with a more realistic optimization, which removes unnecessary $X$ gates from a unitary program.
Finally, we verify a transformation that turns arbitrary \sqire programs into \sqire programs that can run on a linear nearest neighbor architecture. 
 
\subsection{Equivalence of \sqire Programs} \label{sec:sqire-properties}

In general, we will be interested in proving that a transformation is \emph{semantics-preserving}, meaning that the transformation does not change the denotation of the program.
When a transformation is semantics-preserving, we say that it is \emph{sound}.
We will express soundness by requiring equivalence between the input and output of the transformation function.
Equivalence over (unitary) \sqire programs is defined as follows:
\begin{coq}
Definition uc_equiv (c1 c2 : ucom) := forall dim, [[c1]]${}_u^{dim}$ = [[c2]]${}_u^{dim}$.
Infix "$\equiv$" := uc_equiv.
\end{coq}
This definition has several nice properties, including the following.
\begin{coq}
Lemma useq_assoc : forall c1 c2 c3, ((c1 ; c2) ; c3) $\equiv$ (c1 ; (c2 ; c3)).

Lemma useq_congruence : forall c1 c1' c2 c2',
    c1 $\equiv$ c1' ->
    c2 $\equiv$ c2' ->
    c1 ; c2 $\equiv$ c1' ; c2'.
\end{coq}
Associativity and congruence are both important for proving soundness of transformations.
For example, in order to prove that the not propagation optimization (which cancels adjacent $X$ gates) is sound,
we need to prove that the program \coqe{c} has the same denotation as \coqe{X q; X q; c} (for any \coqe{c}). We reason as follows: \coqe{X q; X q; c $\equiv$ (X q; X q); c} by associativity. \coqe{(X q; X q); c $\equiv$ uskip; c} by applying congruence and using the fact that \coqe{X q; X q $\equiv$ uskip}. Finally, \coqe{uskip; c $\equiv$ c} by the identity that says that we can remove a skip on the left without affecting a program's denotation.
 
\subsection{Skip Removal}

\begin{figure}
    \begin{coq}
    Fixpoint rm_uskips (c : ucom) : ucom :=
      match c with
      | c1 ; c2 => match rm_uskips c1, rm_uskips c2 with
                  | uskip, c2' => c2'
                  | c1', uskip => c1'
                  | c1', c2'   => c1'; c2'
                  end
      | c'      => c'
      end.
    \end{coq}
    \caption{Skip removal optimization.}
    \label{fig:rm-skips}
\end{figure}

The skip removal function is shown in Figure~\ref{fig:rm-skips}. 
To show that this function is semantics-preserving, we prove the following lemma.
\begin{coq}
Lemma rm_uskips_sound : forall c, c $\equiv$ (rm_uskips c).
\end{coq}
The proof is straightforward and relies on the identities \texttt{uskip; c $\equiv$ c} and \texttt{c; uskip $\equiv$ c} (which are also easily proven in our development).

We can also prove other useful structural properties about \texttt{rm\_uskips}. 
For example, we can prove that the output of \texttt{rm\_uskips} is either a single skip operation, or contains no skip operations.
\begin{coq}
Inductive skip_free : ucom -> Prop :=
  | SF_seq : forall c1 c2, skip_free c1 -> skip_free c2 -> skip_free (c1; c2)
  | SF_app : forall n l (u : Unitary n), skip_free (uapp u l).

Lemma rm_uskips_correct : forall c,
    (rm_uskips c) = uskip \/ skip_free (rm_uskips c).
\end{coq}
We can also prove that the output of \texttt{rm\_uskips} contains no more skip operations or unitary applications that the original input program.
\begin{coq}
Fixpoint count_ops (c : ucom) : nat :=
  match c with
  | c1; c2 => (count_ops c1) + (count_ops c2)
  | _ => 1
  end.

Lemma rm_uskips_reduces_count : forall c, count_ops (rm_uskips c) <= count_ops c.
\end{coq}

\subsection{Not Propagation}

We now present a more realistic optimization, which removes unnecessary $X$ gates from a program.
This optimization is used as a pre-processing step in a recent quantum circuit optimizer \cite{Nam2018}.
For each $X$ gate in the circuit, this optimization will propagate the gate as far right as possible, commuting through the target of $CNOT$ gates, until a cancelling $X$ gate is found. 
If a cancelling $X$ gate is found, then both gates are removed from the circuit.
If no cancelling $X$ gate is found, then the propagated gate is returned to its original position. 
The structure of this optimization function, and its associated proofs, can be adapted to other propagation-based optimizations (e.g. the ``single-qubit gate cancellation'' routine from the same optimizer \cite{Nam2018}). 

We have proven that this optimization is semantics-preserving.
The main lemmas that the proof relies on are the following.
\begin{coq} 
Lemma XX_id : forall q, uskip ≡ X q; X q. 

Lemma X_CNOT_comm : forall c t, X t; CNOT c t ≡ CNOT c t ; X t.

Lemma U_V_comm : forall (m n : nat) (U V : Unitary 1),
  m <> n -> (U m ; V n) ≡ (V n ; U m). 

Lemma U_CNOT_comm : forall (q n1 n2 : nat) (U : Unitary 1),
  q <> n1 -> q <> n2 -> (U q ; CNOT n1 n2) ≡ (CNOT n1 n2 ; U q). 
\end{coq}
The first lemma says that adjacent $X$ gates cancel. 
The second lemma says that an $X$ gate commutes through the target of a $CNOT$.
The third and fourth lemmas says that single-qubit unitary $U$ commutes with any other 1- or 2-qubit unitary that accesses distinct qubits.
This final lemma is necessitated by our representation of circuits as a list of instructions: In order to discover adjacent $X$ gates, we may need to superficially reorder the instruction list.

\subsection{Circuit Mapping}

We can also use \sqire to verify another useful class of program transformations---mapping algorithms.
Similar to how optimization aims to reduce qubit and gate usage to make programs more feasible to run on near-term machines, circuit mapping aims to address the connectivity constraints of near-term machines \cite{Saeedi2011, Zulehner2017}.
Circuit mapping algorithms take as input an arbitrary circuit and output a circuit that respects the connectivity constraints of some underlying architecture.
To our knowledge, no previous circuit mapping algorithm has been developed with verification in mind.

Here we consider a toy architecture and mapping algorithm.
We assume a linear nearest neighbor (LNN) architecture where qubits are connected to adjacent qubits in the global register (so qubit $i$ is connected to qubits $i-1$ and $i+1$, but qubit 0 and qubit $dim - 1$ are not connected). 
A program will be able to run on our LNN architecture if all $CNOT$ operations occur between connected qubits.
We can represent this constraint as follows.
\begin{coq} 
Inductive respects_LNN : ucom -> Prop :=
  | LNN_skip : respects_LNN uskip
  | LNN_seq : forall c1 c2, 
      respects_LNN c1 -> respects_LNN c2 -> respects_LNN (c1; c2)
  | LNN_app_u : forall (U : Unitary 1) q, respects_LNN (U q)
  | LNN_app_cnot_left : forall n, respects_LNN (CNOT n (n+1))
  | LNN_app_cnot_right : forall n, respects_LNN (CNOT (n+1) n).
\end{coq}
This definition says that skip and single-qubit unitary operations always satisfy the LNN constraint, a sequence construct satisfies the LNN constraint if both of its components do, and a $CNOT$ satisfies the LNN constraint if its arguments are adjacent in the global register.

We map a program to this architecture by adding SWAP operations before and after every $CNOT$ so that the target and control are adjacent when the $CNOT$ is performed, and are returned to their original positions before the next operation.
This algorithm inserts many more SWAPs than the optimal solution, but our verification framework could be applied to optimized implementations as well.

We have proven that this transformation is sound, and that the output program satisfies the LNN constraint.

\section{\sqire for General Verification}
\label{sec:general-verification}

\sqire is useful for more than just verifying program optimizations.
Its simple structure and semantics also allow us to easily verify general properties of quantum programs. 
This makes \sqire a useful tool for reasoning about correctness of low-level quantum programs and thus, we believe, a good candidate for introducing students to concepts of verification and quantum computing.

In this section we discuss correctness properties of three quantum programs, written in \sqire, that could be introduced in an introductory course on quantum computing.

\subsection{GHZ State Preparation}
The Greenberger-Horne-Zeilinger (GHZ) state \cite{Greenberger1989} is an $n$-qubit entangled quantum state of the form
\begin{align*}
    \ket{\text{GHZ}} = \frac{1}{\sqrt{2}}(\ket{0}^{\otimes n}+\ket{1}^{\otimes n}).
\end{align*}
This vector can be defined in Coq as follows:
\begin{coq}
Definition ghz (n : nat) : Matrix (2 ^ n) 1 :=
  match n with 
  | 0 => I 1 
  | S n' => $1/\sqrt{2}$ .* (nket n $\ket{0}$) .+ $1/ \sqrt{2}$ .* (nket n $\ket{1}$)
  end.
\end{coq}
Above, \texttt{nket n $\ket{i}$} is the tensor product of $n$ copies of the basis vector $\ket{i}$.
The GHZ state can be prepared by a circuit that begins with all qubits initialized to the $\ket{0}$ state, prepares a $\ket{+}$ state in the first qubit, and then applies a $CNOT$ to every other qubit with the previous qubit as the control.
A circuit that prepares the 3-qubit GHZ state is shown below, on the right.
The \sqire description of (the unitary portion of) this circuit can be produced by the recursive function below, on the left.

\begin{minipage}{0.55\textwidth}
\begin{coq}
Fixpoint GHZ (n : nat) : ucom :=
  match n with
  | 0 => uskip
  | 1 => H 0
  | S n' => GHZ n'; CNOT (n'-1) n'
  end.
\end{coq}
\end{minipage}
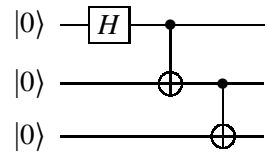
\begin{minipage}{0.4\textwidth}
\[
  \Qcircuit @C=1em @R=1em {
    \lstick{\ket{0}} & \gate{H} & \ctrl{1} & \qw & \qw \\
    \lstick{\ket{0}} & \qw & \targ & \ctrl{1} & \qw \\
    \lstick{\ket{0}} & \qw & \qw & \targ & \qw
    }
\]
\end{minipage}

The function \coqe{GHZ} describes a \emph{family} of \sqire circuits: For every $n$, \coqe{GHZ n} is a valid \sqire program and quantum circuit. 
We aim to show via an inductive proof that every circuit generated by \coqe{GHZ n} produces the corresponding \coqe{ghz n} vector when applied to $\ket{0\dots0}$.
We prove the following theorem:
\begin{coq}
Theorem ghz_correct : forall n : nat, [[GHZ n]]${}_u^n$ $\times$ nket n $\ket{0}$ = ghz n.
\end{coq}
The proof applies induction on $n$. 
For the base case, we show $H$ applied to $\ket{0}$ produces the $\ket{+}$ state.
For the inductive step, the induction hypothesis says that the result of applying \texttt{GHZ n'} to the input state \texttt{nket n $\ket{0}$} produces the state
\begin{coq}
$1/\sqrt{2}$ .* (nket n' $\ket{0}$) .+ $1/\sqrt{2}$ .* (nket n' $\ket{1}$) $\otimes$  $\ket{0}$.
\end{coq}
By considering the effect of applying \texttt{CNOT (n'-1) n'} to this state, we can complete the proof.

\subsection{Teleportation}

Quantum teleportation is one of the first quantum programs shown in introductory classes on quantum computing.
We present it here to highlight differences between the density matrix semantics and non-deterministic semantics presented in Section~\ref{sec:general-sqire}. The \sqire program for teleportation is given below.
\begin{coq}
Definition bell : com := H 1 ; CNOT 1 2.
Definition alice : com := CNOT 0 1 ; H 0 ; meas 0 ; meas 1.
Definition bob : com := CNOT 1 2; CZ 0 2; reset 0; reset 1.
Definition teleport : com := bell; alice; bob.
\end{coq}


The correctness property of quantum teleportation says that the input qubit is the same as the output qubit. Since \sqire does not permit us to discard qubits, we instead reset them to $\ket{0}$.
Under the density matrix semantics, we aim to prove the following:
\begin{coq}
Lemma teleport_correct : forall (ρ : Density (2^1)),
  WF_Matrix ρ -> 
  [[teleport]]${}_d^3$ (ρ ⊗ ∣0⟩⟨0∣ ⊗ ∣0⟩⟨0∣) = ∣0⟩⟨0∣ ⊗ ∣0⟩⟨0∣ ⊗ ρ.
\end{coq}

The proof for the density matrix semantics is simple: We compute the products with our matrix solver, and do some simple (automated) arithmetic to show that the output matrix has the desired form. While short, this proof does not give much intuition about how quantum teleportation works.

For the non-deterministic semantics the proof is more involved, but also more illustrative of the inner workings of the teleport algorithm.
Under the non-deterministic semantics, we aim to prove the following:
\begin{coq}
Lemma teleport_correct : forall (ψ : Vector (2^1)) (ψ' : Vector (2^3)),
  WF_Matrix ψ ->
  teleport / (ψ  ⊗ ∣ 0 , 0 ⟩) ⇩ ψ' ->
  ψ' ∝ ∣ 0 , 0 ⟩ ⊗ ψ.   
\end{coq}

Since the non-deterministic semantics does not rescale outcomes, we merely require that every outcome is proportional to ($\propto$) the intended outcome. Note that this statement is quantified over every outcome $\psi'$ and hence all possible paths to $\psi'$. If instead we simply claimed that 
\begin{center}
\begin{tabular}{c}
    \begin{coq}
    teleport / (ψ  ⊗ ∣ 0 , 0 ⟩) ⇩ ∣0 , 0 ⟩ ⊗ ψ
    \end{coq} 
\end{tabular}
\end{center}
we would only be stating that some such path exists.

The first half of the circuit is unitary, so we can simply compute the effect of applying a $H$ gate, two $CNOT$s and another $H$ gate to the input state. We can then take both measurement steps, leaving us with four different cases to prove correct. In each of the four cases, we can use the outcomes of measurement to correct the final qubit, putting it into the state $\psi$. Finally, resetting the already-measured qubits is deterministic, and leaves us in the desired state.

\subsection{The Deutsch-Jozsa Algorithm}

In the quantum query model, we are given access to a Boolean function $f:\{0,1\}^n\to\{0,1\}$ through an oracle defined by the map $U_f:\ket{y,x}\mapsto\ket{y\oplus f(x),x}$.
For a function $f$ on $n$ bits, the unitary matrix $U_f$ is a linear operator over a $2^{n+1}$ dimensional Hilbert space.
In order to describe the Deutsch-Jozsa algorithm in \sqire, we must first give a \sqire definition of oracles.

To begin, note that any $n$-bit Boolean function $f$ can be written as 
\begin{align*}
    f(x_1,\ldots,x_n)
    = \left\{\begin{array}{ll}
    f_0(x_1,\ldots,x_{n-1}) & \text{if } x_n=0 \\
    f_1(x_1,\ldots,x_{n-1}) & \text{if } x_n=1
    \end{array}\right.
\end{align*}
where $f_b(x_1,\ldots,x_{n-1})=f(x_1,\ldots,x_{n-1},b)$ is a Boolean function on $(n-1)$ bits for $b\in\{0,1\}$.
Similarly, an oracle can be written as 
$U_f= U_{f_0}\otimes\ket{0}\bra{0} +  U_{f_1}\otimes\ket{1}\bra{1}$ for $U_f\ket{y,x_1,\ldots,x_{n-1},b}=U_{f_b}\ket{y,x_1,\ldots,x_{n-1}}\ket{b}$. 
In the base case ($n=0$), a Boolean function is a constant function of the form $f(\perp)=0$ or $f(\perp)=1$ and an oracle is either the identity matrix, i.e., $\ket{y}\mapsto\ket{y}$, or a Pauli-$X$ matrix, i.e., $\ket{y}\mapsto\ket{y\oplus 1}$.
As a concrete example, consider the following correspondences between the 1-bit Boolean functions and $4\times 4$ unitary matrices:
\begin{align*}\nonumber
    f_{00}(x)&=0 & U_{f_{00}} &= I\otimes\ket{0}\bra{0} + I\otimes\ket{1}\bra{1}, \\\nonumber
    f_{01}(x)&=1-x & U_{f_{01}} &= X\otimes\ket{0}\bra{0} + I\otimes\ket{1}\bra{1}, \\\nonumber
    f_{10}(x)&=x & U_{f_{10}} &= I\otimes\ket{0}\bra{0} + X\otimes\ket{1}\bra{1}, \\\label{eq:deutsch-1}
    f_{11}(x)&=1 & U_{f_{11}} &= X\otimes\ket{0}\bra{0} + X\otimes\ket{1}\bra{1}.
\end{align*}
The observation above enables the following inductive definition of an oracle.
\begin{coq}
Inductive boolean : nat -> ucom -> Set :=
  | boolean_I : forall u, u ≡ uskip -> boolean 0 u
  | boolean_X : forall u, u ≡ X 0 -> boolean 0 u
  | boolean_U : forall u u1 u2 dim,
                boolean dim u1 -> 
                boolean dim u2 ->
                [[u]]${}_u^{2 + dim}$ = [[u1]]${}_u^{1 + dim}$ ⊗ $\ket{0}\bra{0}$ .+ [[u2]]${}_u^{1 + dim}$ ⊗ $\ket{1}\bra{1}$ ->
                boolean (1 + dim) u.
\end{coq}
\coqe{boolean dim U} describes an oracle for a \coqe{dim}-bit Boolean function
whose denotation is a $2^{1+dim} \times 2^{1+dim}$ unitary matrix.

A Boolean function is balanced if the number of inputs that evaluate to $1$ is exactly half of the domain size. 
A Boolean function is constant if for all inputs, the function evaluates to the same output, i.e., $\forall x.~f(x)=0$ or $\forall x.~f(x)=1$.
Given an oracle, we can determine whether it describes a balanced or constant function by counting the number of inputs that evaluate to 1.
We define balanced and constant oracles in \sqire as follows.
\begin{coq}
Fixpoint count {dim : nat} {U : ucom} (P : boolean dim U) : C :=
  match P with
  | boolean_I _ _ => 0
  | boolean_X _ _ => 1
  | boolean_U _ _ _ _ P1 P2 _ => count P1 + count P2
  end.
Definition balanced {dim : nat} {U : ucom} (P : boolean dim U) : Prop :=
  dim >= 1 /\ count P = 2 ^ (dim - 1).
Definition constant {dim : nat} {U : ucom} (P : boolean dim U) : Prop :=
  count P = 0 \/ count P = 2 ^ dim.
\end{coq}

\begin{figure}[t]
\begin{subfigure}{0.65\textwidth}
\begin{coq}
Fixpoint cpar (n : nat) (u : nat -> ucom) :=
  match n with
  | 0 => uskip
  | S n' => cpar n' u ; u n'
  end.
Definition deutsch_jozsa (n : nat) (U : ucom) :=
  X 0 ; cpar n H ; U ; cpar n H.
\end{coq}
\end{subfigure}
\begin{subfigure}{0.34\textwidth}
\centerline{
  \Qcircuit @C=1em @R=1em {
     \lstick{\ket{0}} & \gate{X} & \gate{H} & \multigate{3}{U} & \gate{H} & \qw & \qw \\
     \lstick{\ket{0}} & \qw & \gate{H} & \ghost{U} & \gate{H} & \meter & \cw \\
     \lstick{\vdots} &  &  &  &  & \vdots &  \\
     \lstick{\ket{0}} & \qw & \gate{H} & \ghost{U} & \gate{H} & \meter & \cw \\
    }
}  
\end{subfigure}
  \caption{The Deutsch-Jozsa algorithm in \sqire and as a circuit.}
  \label{fig:deutsch}
\end{figure}
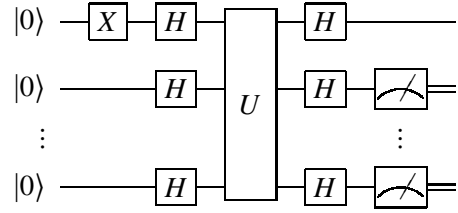

In the Deutsch-Jozsa problem \cite{deutsch1992rapid}, we are promised that the function $f$ is either balanced or constant, and the goal is to decide which is the case by querying the oracle.
The Deutsch-Jozsa algorithm begins with an all-$\ket{0}$ state, and prepares the input state \coqe{∣-⟩ ⊗ nket dim ∣+)}. This state is prepared by applying an $X$ gate on the first qubit, and then applying a $H$ gate to every qubit in the program.
Next, the oracle $U$ is queried, and a $H$ gate is again applied to every qubit in the program.
Finally, all qubits except the first are measured in the standard basis.
This algorithm is shown as a circuit and in \sqire in \cref{fig:deutsch}.
Note the use of Coq function \coqe{cpar}, which constructs a \sqire program that applies the same operation to every qubit in the program.

If the outcome of measurement is an all-zero string, then the algorithm outputs ``accept,'' which indicates that the function is constant.
Otherwise the algorithm outputs ``reject''.
Formally, the algorithm will output ``accept'' when the output state is supported on $\Pi = I\otimes \ket{0}\bra{0}^{\otimes dim}$ and output ``reject'' when the output state is orthogonal to $\Pi$.
We can express this in Coq as follows.
\begin{coq}
Definition accept {dim : nat} {U : ucom} (P : boolean dim U) : Prop :=
  exists (ψ : Matrix 2 1), 
  ((ψ ⊗ nket dim ∣0⟩)† × [[deutsch_jozsa (1+dim) U]]${}_u^{1 + dim}$ × (nket (1+dim) ∣0⟩))) 0 0 = 1. 
Definition reject {dim : nat} {U : ucom} (P : boolean dim U) : Prop :=
  forall (ψ : Matrix 2 1), WF_Matrix ψ -> 
  ((ψ ⊗ nket dim ∣0⟩)† × [[deutsch_jozsa (1+dim) U]]${}_u^{1 + dim}$ × (nket (1+dim) ∣0⟩))) 0 0 = 0. 
\end{coq}
We now prove the following theorems.
\begin{coq}
Theorem deutsch_jozsa_constant_correct :
  forall (dim : nat) (U : ucom) (P : boolean dim U), constant P -> accept P.
Theorem deutsch_jozsa_balanced_correct :
  forall (dim : nat) (U : ucom) (P : boolean dim U), balanced P -> reject P.
\end{coq}

The key lemma in our proof states that the probability of outputting ``accept'' depends on the number of inputs that evaluate to 1, i.e., \coqe{count P}.
\begin{coq}
Lemma deutsch_jozsa_success_probability :
  forall {dim : nat} {U : ucom} (P : boolean dim U) (ψ : Matrix 2 1) (WF : WF_Matrix ψ),
    (ψ ⊗ nket dim ∣0⟩)† × [[deutsch_jozsa (1 + dim) U]]${}_u^{1 + dim}$ × (nket (1 + dim) ∣0⟩))
    = (1 - 2 * count P * /2 ^ dim) .* (ψ† × ∣1⟩).
\end{coq}
This lemma is proved by induction on \coqe{P}, which is the proof that \coqe{U} is an oracle.
We sketch the structure of the proof below, using mathematical notation for ease of presentation.

In the base case, either \coqe{U ≡ uskip} or \coqe{U ≡ X 0}.
Observing that $\bra{\psi}H X^bH\ket{1}=\left(1-2b\right)\langle\psi|1\rangle$, we can complete the proof by direct calculation on matrices.
For the inductive step,
the induction hypothesis says that, for any Boolean function $g$ of $dim$ bits,
\[\bra{\psi,0^{dim}}~H^{\otimes (1+dim)}U_{g}H^{\otimes (1+dim)}~\ket{1,0^{dim}}=\left(1 - \frac{2 |S(g)|}{2^{dim}}\right)\langle\psi|1\rangle,\] 
where $|S(g)|$ is the number of inputs on which $g$ evaluates to 1. 
Therefore, for $(1+dim)$-bit Boolean function $f$,
since $|S(f)|=|S(f_0)|+|S(f_1)|$,
\begin{align*}
    & \bra{\psi,0^{1+dim}}~H^{\otimes (2+dim)}U_f H^{\otimes (2+dim)}~\ket{1,0^{1+dim}} \\
    &\qquad = 
    \bra{\psi,0^{1+dim}}~H^{\otimes (2+dim)} (U_{f_0}\otimes\ket{0}\bra{0}
    + U_{f_1}\otimes\ket{1}\bra{1})H^{\otimes (2+dim)}~\ket{1,0^{1+dim}} \\
    &\qquad = 
    \frac{1}{2}\bra{\psi,0^{dim}}~ H^{\otimes (1+dim)}U_{f_0}H^{\otimes (1+dim)}~\ket{1,0^{dim}}
    + \frac{1}{2}\bra{\psi,0^{dim}}~H^{\otimes (1+dim)}U_{f_1}H^{\otimes (1+dim)}~\ket{1,0^{dim}} \\
    &\qquad 
    =\left(1 - \frac{2|S(f)|}{2^{1+dim}}\right)\langle\psi|1\rangle.
\end{align*}

\section{Conclusions and Future Work}

We have presented \sqire, a simple, low-level quantum language embedded in Coq.
We argued that \sqire can serve as an intermediate representation for compiled quantum programs and verified several transformations of \sqire programs.
Previous work has considered the problem of verified compilation of Boolean circuits \cite{amy18reversible, Rand2018} and verified optimization of ZX terms \cite{Fagan2018}, but, to our knowledge, our \sqire-based transformations are the first verified optimizations for a realistic low-level quantum language.

We also showed how to directly verify the correctness of \sqire programs.
Compared to previous languages for verification of quantum programs, \sqire is easy to learn and straightforward to use and thus, we believe, a good candidate for pedagogy. We hope that \sqire can be used as the basis for an introduction to quantum computing in the style of Software Foundations \cite{Pierce2016}. 

Moving forward, we plan to make more progress toward a full-featured verified compilation stack for quantum programs, from verified transformation of high-level quantum languages to \sqire code to verified production of circuits that run on real quantum machines, following the vision of a recent Computing Community Consortium report \cite{Martonosi2019}. We also plan to implement additional verified mapping and optimization functions, and transformation functions that take into account other limitations of near-term quantum machines, such as their high rate of error, as envisioned by Rand et al. \cite{Rand2019}. 
We are also looking at extending \sqire with branching measurements and while loops, in the style of Selinger's QPL \cite{Selinger2004} or Ying's quantum while language \cite{Ying2011}, allowing us to implement and verify the many recently-developed quantum Hoare logics \cite{Hung2019, Unruh2019a, Unruh2019, Ying2011}. 


\section*{Acknowledgments}

We gratefully acknowledge the support of the U.S. Department of Energy, Office of Science, Office of Advanced Scientific Computing Research, Quantum Testbed Pathfinder Program under Award Number DE-SC0019040.

\bibliographystyle{eptcs}
\bibliography{references}

\appendix

\section{\qwire vs. \sqire}
\label{app:comparison}

The fundamental difference between \sqire and its sibling, \qwire \cite{Paykin2017}, is that \sqire relies on a global register of qubits and every operation is applied to an explicit set of qubits within the global register. By contrast, \qwire uses Higher Order Abstract Syntax \cite{Pfenning1988} to take advantage of Coq's variable binding and function composition facilities. \qwire circuits have the following form:
\begin{coq}
Inductive Circuit (w : WType) : Set :=
| output : Pat w -> Circuit w
| gate   : forall {w1 w2}, 
           Gate w1 w2 ->  Pat w1 -> (Pat w2 -> Circuit w) -> Circuit w
| lift   : Pat Bit -> (bool -> Circuit w) -> Circuit w.
\end{coq}

Patterns \coqe{Pat} type the variables in \qwire circuits and have a specific wire type \coqe{w}, corresponding to some collection of bits and qubits. In the definition of \coqe{gate}, we provide a parameterized \coqe{Gate}, an appropriate input pattern, and a \emph{continuation} of the form \coqe{Pat w2 -> Circuit w}, which is a placeholder for the next gate to connect to. This is evident in the definition of the composition function:  
\begin{coq}
Fixpoint compose {w1 w2} (c : Circuit w1) (f : Pat w1 -> Circuit w2) : Circuit w2 :=
  match c with 
  | output p     => f p
  | gate g p c'  => gate g p (fun p' => compose (c' p') f)
  | lift p c'    => lift p (fun bs => compose (c' bs) f)
  end.
\end{coq}
In the \coqe{gate} case, the continuation is applied directly to the output of the first circuit.

Circuits correspond to open terms; closed terms are represented by \emph{boxed} circuits:
\begin{coq}
Inductive Box w1 w2 : Set := box : (Pat w1 -> Circuit w2) -> Box w1 w2.
\end{coq}
This representation allows for easy composition: Any two circuits with matching input and output types can easily be combined using standard function application. For example, consider the following convenient functions for sequential and parallel composition of closed terms:
\begin{coq}
Definition inSeq {w1 w2 w3} (c1 : Box w1 w2) (c2 : Box w2 w3): Box w1 w3 :=
  box p1 ⇒ 
    let p2 ← unbox c1 p1;
    unbox c2 p2.

Definition inPar {w1 w2 w1' w2'} 
                 (c1 : Box w1 w2) (c2 : Box w1' w2') : Box (w1 ⊗ w1') (w2 ⊗ w2'):=
  box (p1,p2) ⇒ 
    let p1'     ← unbox c1 p1;
    let p2'     ← unbox c2 p2; 
    (p1',p2').
\end{coq}

Unfortunately, proving useful specifications for these functions 
is quite difficult. Since the denotation of a circuit must be (in the unitary case) a square matrix of size $2^n$ for some $n$, we need to map all of our variables to $0$ through $n - 1$, ensuring that the mapping function has no gaps even when we initialize or discard qubits. We maintain this invariant through compiling to a de Bruijin-style variable representation~\cite{deBruijn1972}. Reasoning about the denotation of our circuits, then, involves reasoning about this compilation procedure. In the case of open circuits (our most basic circuit type), we must also reason about the contexts that type the available variables, which change upon every gate application. 

As informal evidence of the difficulties of \qwire's representation on proof, we note that while the proof of \coqe{inPar}'s correctness in \sqire (see \cref{app:sqire-composition}) took a matter of hours, we still lack a correctness proof for the corresponding function in \qwire after many months of trying.

Of course, this comparison is not entirely fair: \qwire's \coqe{inPar} is more powerful than \sqire's equivalent. \sqire's \coqe{inPar} function does not require every qubit within the global register to be used -- any gaps will be filled by identity matrices. Also, \sqire does not allow introducing or discarding qubits, which we suspect will make ancilla management difficult. 

Another important difference between \qwire and \sqire is that \qwire circuits cannot be easily decomposed into smaller circuits because output variables are bound in different places in the circuit. By contrast, a \sqire program is an arbitrary nesting of smaller programs, and \coqe{c1;((c2;(c3;c4));c5)} is equivalent to \coqe{c1;c2;c3;c4;c5} under all semantics, whereas every \qwire circuit (only) associates to the right. As such, rewriting using \sqire identities is substantially easier.

There are other noteworthy difference between the two languages. \qwire's standard denotation function is in terms of superoperators over density matrices, which are harder to work with than simple unitary matrices. \qwire also provides additional useful tools for quantum programming, like wire types and support for \emph{dynamic lifting}, which passes the control flow to a classical computer before resuming a quantum computation. 

The differences between these tools stem from the fact that \qwire was developed as a programming language for quantum computers~\cite{Paykin2017}, and was later used as a verification tool~\cite{Rand2017, Rand2018}. By contrast, \sqire is mainly a tool for verifying quantum programs, ideally compiled from another language such as Q\#~\cite{Svore2018}, Quipper~\cite{Green2013} or \qwire itself.

\section{Composition in \sqire} \label{app:sqire-composition}

\sqire was not designed to be compositional.
As such, describing the composition of \sqire programs can be difficult.
To begin, consider the following function, which composes two \sqire programs in parallel.
\begin{coq}
Fixpoint map_qubits (f : nat -> nat) (c : ucom) :=
  match c with
  | uskip => uskip
  | c1; c2 => map_qubits f c1; map_qubits f c2
  | uapp u l => uapp u (map f l)
  end.

Definition inPar (c1 c2 : ucom) (d1 : nat) :=
  c1; map_qubits (fun q => q + d1) c2.
\end{coq}
The correctness property for \coqe{inPar} says that the denotation of \coqe{inPar c1 c2} can be constructed from the denotations of \coqe{c1} and \coqe{c2}.
\begin{coq}
Lemma inPar_correct : forall c1 c2 d1 d2,
  uc_well_typed d1 c1 ->
  [[inPar c1 c2 d1]]${}_u^{d1 + d2}$ = [[c1]]${}_u^{d1}$ ⊗ [[c2]]${}_u^{d2}$.
\end{coq}
The \coqe{inPar} function is relatively simple, but more involved than the corresponding \qwire definition because it requires relabeling the qubits in program $c_2$.

General composition in \sqire requires more involved relabeling functions that are less straightforward to describe.
For example, consider the composition expressed in the following \qwire program:
\begin{coq}
box (ps, q) =>
  let (x, y, z) <- unbox c1 ps;
  let (q, z) <- unbox c2 (q, z);
  (x, y, z, q).
\end{coq}
This program connects the last output of program $c_1$ to the second input of program $c_2$. This operation is natural in \qwire, but describing this type of composition in \sqire requires some effort. In particular, the programmer must determine the required size of the new global register (in this case, 4) and explicitly provide a mapping from qubits in $c_1$ and $c_2$ to indices in the new register (for example, the first qubit in $c_2$ might be mapped to the fourth qubit in the new global register). 
When \sqire programs are written directly, this puts extra burden on the programmer.
When \sqire is used as an intermediate representation, however, these mapping functions should be produced automatically by the compiler.
The issue remains, though, that any proofs we write about the result of composing $c_1$ and $c_2$ will need to reason about the mapping function used (whether produced manually or automatically).

\end{document}